\documentclass[twocolumn,showpacs,preprintnumbers,amsmath,amssymb,10pt,a4paper,showkeys]{revtex4}

\usepackage{geometry}
\usepackage{graphicx}
\usepackage{dcolumn}
\usepackage{bm}
\newcommand{\be}{\begin{equation}}
\newcommand{\ee}{\end{equation}}
\newcommand{\bd}{\begin{displaymath}}
\newcommand{\ed}{\end{displaymath}}
\newcommand{\BE}{\begin{eqnarray}}
\newcommand{\EE}{\end{eqnarray}}

\newcommand{\bx}{\ensuremath{\mathbf{x}}}
\newcommand{\by}{\ensuremath{\mathbf{y}}}

\newcommand{\bn}{\ensuremath{\mathbf{n}}}

\newcommand{\bee}{\ensuremath{\mathbf{e}}}

\newcommand{\boldeff}{\ensuremath{\mathbf{f}}}

\newcommand{\boldeta}{{\mbox{\boldmath $\eta$}}}

\newcommand{\bxi}{\bm{\xi}}

\newcommand{\avg}[1]{\left\langle{#1}\right\rangle}

\begin{document}

\preprint{}
\title{Independence and interdependence in the nest-site choice by honeybee swarms:\\
agent-based models, analytical approaches and pattern formation}

\author{Tobias Galla}
\email{tobias.galla@manchester.ac.uk}

\affiliation{Theoretical Physics, School of Physics and Astronomy, The University of Manchester, Manchester M13 9PL, United Kingdom}

\date{\today}

\begin{abstract}
In a recent paper List, Elsholtz and Seeley \cite{list} have devised an
agent-based model of the the nest-choice dynamics in swarms of
honeybees, and have concluded that both interdependence and
independence are needed for the bees to reach a consensus on the best
nest site. We here present a simplified version of the model which can be treated
analytically with the tools of statistical physics and which largely
has the same features as the original dynamics. Based on our
analytical approaches it is possible to characterize the co-ordination
outcome exactly on the deterministic level, and to a good
approximation if stochastic effects are taken into account, reducing
the need for computer simulations on the agent-based level. In the
second part of the paper we present a spatial extension, and show that
transient non-trivial patterns emerge, before consensus is
reached. Approaches in terms of Langevin equations for continuous
field variables are discussed.
\end{abstract}

\maketitle
\section{Introduction}
In a recent paper \cite{list} List, Elsholtz and Seeley (LES for short
from now on) presented an agent-based model of the decision making
processes in the nest-site choice of honeybee swarms, and their paper
has attracted a significant amount of attention in the public media
(see e.g. The Economist of 13 February 2009). LES focus on the
interplay between the interdependence among bees and their
independence in assessing the quality of potential nest sites, and
conclude that, assuming a degree of independence, bees will generally
choose the best nest site for a wide range of non-extremal model
parameters, and secondly that both independence and interdependence
are necessary for the swarm to reach a reliable consensus on the best
possible nest site. The precise notions of `independence' and
`interdependence' will become clear below, but in essence LES find
that, in order to identify the best nest site, the bees in their model need
to make independent assessments of the true quality of nest sites
instead of blindly following the choices advertised by fellow bees. At
the same time interdependence, i.e. direct communication between bees,
is required as well in order to promote a quick and accurate
convergence towards a consensus on the best choice.

While the work of LES is mainly based on computer simulations, we here
discuss a modification of their agent-based dynamics, which can be
addressed using the tools and techniques of statistical
mechanics. Such analytical solutions mean that computer simulations on
the agent-based level could in principle be dispensed with entirely,
although admittedly our resulting equations still require numerical
solution. Still, if there are $k$ nest sites in the model, then on the
deterministic level, the solution of the reduced model comes out as a
coupled set of $k$ quadratic equations for the number of bees
advertising the different sites, which is a significant reduction in
complexity of the problem. LES choose $k=5$ in their analysis, so that
we are able to reduce the deterministic features of their model to a
very small set of equations. Of course, stochasticity is important in
the model, but given the deterministic solution, fluctuations about
the deterministic limit can be characterized analytically as well to a
good approximation, as we will describe below. The main aim of this
current paper is to present the modified version of LES's model, and
to show how it can be addressed with the tools of statistical
physics. In the second part of our work we then extend the model of
\cite{list} to include space, and study the pattern formation dynamics
of this extended model.

The natural process modeled by LES is the following: once a
year once a colony of bees has reached a certain size, a `search
committee' consisting of several hundred bees swarm out to identify
suitable nest sites, and once an agreement has been reached the
original colony divides, and the queen leaves with about two thirds of
the bees for the new nest (see \cite{list} and references therein). The process LES are interested
in is the actual decision making. The bees who fly out to inspect
potential new sites return to the original colony and perform dances
to advertise the new sites they have inspected. The duration of the
dance here is a measure for the quality of a newly inspected site,
the better the site, the longer the dance. Fellow bees who observe
these dances might then themselves start dancing for that particular
site as well, and this may happen with or without an independent
inspection of the actual site, i.e. with or without an independent
assessment of the site's quality. This may then lead to other bees
initiating dances (again with or without independent assessment of the
site's quality), and finally, if all goes well, a general consensus on
one of the nest sites will be reached. The question LES address in
their model is whether this consensus will be on the best possible
site, or whether a sub-optimal choice can result.

In order to model communication and interdependence between bees LES
assume that the rate with which a given bee starts dancing for a
particular site depends on an a-priori rate of independently discovering
that site, and, weighted by an `interdependence coefficient', the
fraction of other bees dancing for this site. If the interdependence
between bees is strong then the latter factor carries a large weight,
and the rate with which bees start dancing for a given site is mostly
determined by the fraction of bees already dancing for that site. If
interdependence is weak dances for all sites are commenced essentially
with equal rates, independently of how many bees are already
advertising the different sites. The second component of the model is
the time a dance lasts once it has been begun.  In the model by LES,
the duration of the dance is determined by a combination of the
actual (or perceived) quality of the site, and a uniform site-independent contribution. In the case of strong independence, the bees are assumed to
inspect the site before dancing for it, and the duration of the dance
will be proportional to the perceived quality of the site. In the case
of weak independent quality assessment the duration of the dance for a given site does
not depend much on the site's quality.

The key question LES address in their study is whether or not the swarm of bees will eventually reach a consensus on what the best nest-site might be, and whether this consensus choice then is the actual best site. The outcome of the decision dynamics here depends on the independence and interdependence parameters introduced above, and, to a lesser extent, on the accuracy with which bees assess the quality of a site if they actually inspect it. I.e. a bee may decide to visit a site before dancing for it, but the perceived quality may be a random function of the actual quality, so that noise may blur the bee's assessment of the site. In this paper we complement the simulation study of LES by analytical computations, allowing for a more extensive characterization of the model. We also present extensions towards a spatial model, and discuss how patterns may emerge at transient times if the model is considered on a two-dimensional regular lattice. The paper is organized as follows: In Sec. \ref{sec:model} we will first re-iterate the definitions of the model by LES, and then introduce the simplified dynamics, which we will then address analytically within a master equation approach in Sec. \ref{sec:ana}. We here formulate deterministic mean-field equations and also address first-order finite-size corrections. In Sec. \ref{sec:space} we then introduce a spatial version of the reduced model, and show in computer simulations that non-trivial transient patterns may emerge if model parameters are chosen suitably. The study of the spatial model is complemented by the numerical integration of a set of partial differential equations for a coarse-grained dynamics. In Sec. \ref{sec:concl} we finally summarize our work and point to future directions which may be of interest.

\section{Model definitions}\label{sec:model}
\subsection{Original model by List et al}

In the original model by LES there are $N$ bees, labelled
$i=1,2,\dots,N$, and $k$ potential nest sites, we label them by
$\alpha=1,\dots,k$. Each nest site has an intrinsic quality $q_\alpha\geq 0$. These `actual'
qualities of the different nest sites are fixed from the beginning and do not change
over time. We always choose $q_1<q_2<\dots<q_k$, i.e. site number $k$ is the best site, and site number $1$ the worst. Additionally, for each potential nest site $\alpha$, there is an a-priori probability $\pi_\alpha$ of finding this particular site. These rates of finding the different nests will be taken to fulfill $\sum_{\alpha=1}^k\pi_\alpha<1$, and we will set $\pi_0=1-\sum_{\alpha=1}^k\pi_\alpha$. Throughout this paper we will follow \cite{list} and assume that the $\pi_\alpha$, $\alpha=1,\dots,k$, are identical for all nests, but an extension of the model to the more general case is straightforward. 

The decision and dancing dynamics within the model then proceeds in discrete time steps,
$t=1,2,3,\dots$, and at each time step each bee can either be dancing
for a particular site $\alpha\in\{1,\dots,k\}$, or not be dancing. The not-dancing state here describes bees that may have flown out to inspect a site, but have not yet started dancing for it, as well as bees who are observing other bees, or bees that are resting \cite{list}. We will sometimes refer to these bees as `passive'. Following LES we write
$s_i(t)=\alpha\in\{1,\dots,k\}$ if bee number $i$ is dancing for site
$\alpha$ at time $t$, and $s_i(t)=0$, if she is not dancing at time
$t$. In LES's model, the state of a bee is characterized by an
additional variable, $d_i(t)\geq 0$, indicating the remaining duration
of bee $i$'s dance at time $t$. I.e., up to rounding to a near integer, if $d_i(t)\geq 1$, then bee $i$ will
be dancing for another $d_i(t)$ time steps. In the non-dancing state (i.e. for a bee $i$ with $s_i(t)=0$) $d_i(t)$ has no relevance. More specifically, the dynamics proceeds
according to the following algorithm:
\begin{enumerate}
\item[1.] At time $t=0$ initialize the $\{s_i(t),d_i(t)\}$, for example set $s_i(t=0)=0$, and $d_i(t)=0$ for all $i$.
\item[2.] At time $t$ compute the fraction of passive bees and of those dancing for the different sites $\{\alpha\}$ at this time, $f_\alpha(t)=N^{-1}\sum_{i=1}^N\delta_{\alpha,s_i(t)}\in[0,1]$ ($\alpha=0,\dots,k$). $\delta$ here stands for the Kronecker delta, i.e. $\delta_{\alpha\beta}=1$ if $\alpha=\beta$, and $\delta_{\alpha\beta}=0$ otherwise. Then compute $p_\alpha(t+1)=(1-\lambda)\pi_\alpha+\lambda f_\alpha(t)$ ($\alpha=0,\dots,k$). Note that $\sum_{\alpha=0}^k p_\alpha(t+1)=1$ by construction, due to the normalization of the $\pi_\alpha$ and of the $f_\alpha$. The variable $\lambda\in[0,1]$ is a model parameter, and its meaning will be explained below.
\item[3.] In order to update the states of the bees distinguish between bees that are not dancing at time $t$ and those which are dancing at time $t$. Update all bees in parallel dynamics, i.e. for all $i=1,\dots,N$ perform the following:
\begin{itemize}
\item[3a)] If bee $i$ is dancing at time $t$, then check whether $d_i(t)\geq1$. If this is the case then the bee keeps dancing for the site she is currently dancing for, i.e. set $s_i(t+1)=s_i(t)$ and reduce the remaining duration of the dance by one unit: $d_i(t+1)=d_i(t)-1$. If the bee is dancing, but $d_i(t)<1$, then set $s_i(t+1)=0$ and $d_i(t+1)=0$ (the bee stops dancing).
\item[3b)] If, at time $t$, bee $i$ is not dancing ($s_i(t)=0$), then with probability $p_\alpha(t+1)$ set $s_i(t+1)=\alpha$ ($\alpha=1,\dots,k$), i.e. with probability $p_1(t+1)$ the bee starts dancing for site $\alpha=1$, with probability $p_2(t+1)$ she starts dancing for site $\alpha=2$ and so on. With probability $p_0(t+1)$ she will remain in the non-dancing state. If a dance is commenced then it remains to specify the duration of the dance bee $i$ has just started for site $\alpha$. Here draw a random number $T$ from a standard Gaussian distribution (mean zero, unit variance). Then set
\begin{displaymath}
~~~~~~~~~~~~~~~~~~~~~d_i(t+1)=\left\{\begin{array}{cl} q_\alpha \exp(\sigma T) & ~\mbox{with prob. } 1-\mu \\ {}&{}\\ K\exp(\sigma T) & ~\mbox{with prob. } \mu \end{array}\right.
\end{displaymath}
\end{itemize}
\item[4.] Iterate, i.e. go to 2.
\end{enumerate}
LES here set $K=q_k$, i.e. $K$ is the quality of the best site. The variables $\lambda\in[0,1]$ and $1-\mu\in[0,1]$ are the interdependence and independence parameters mentioned in the introduction. If $\lambda$ is close to one, then the interdependence between bees is strong, i.e. the rate of commencing a dance for a given site is mostly determined by the number of bees already dancing for this site. At small $\lambda$ the dancing activity of other bees is mostly irrelevant for a given individual to take up a dance. The independence in the assessment of the nest-site qualities is parametrized by $\mu$, more accurately by $1-\mu$ in the notation of LES which we will follow here. For large $1-\mu$ the duration of a dance for site $\alpha$ strongly depends on the quality $q_\alpha$ of that site (independent assessment), whereas for small $1-\mu$, the actual quality of the different sites is mostly irrelevant for the duration of the dances. The variable $\sigma\geq 0$ is a further model parameter and represents the reliability of the bees' assessment of the qualities of the different sites. As seen in the above update rules the perceived quality of site $\alpha$ is given by $q_\alpha \exp(\sigma T)$, where $T$ is a standard Gaussian random variable. Thus if $\sigma=0$, the perceived quality is always identical to the actual quality $q_\alpha$. Increasing $\sigma$ introduces more and more uncertainty into this perception.

\begin{figure}[t]
\vspace{2em}
\centerline{\includegraphics[width=0.4\textwidth]{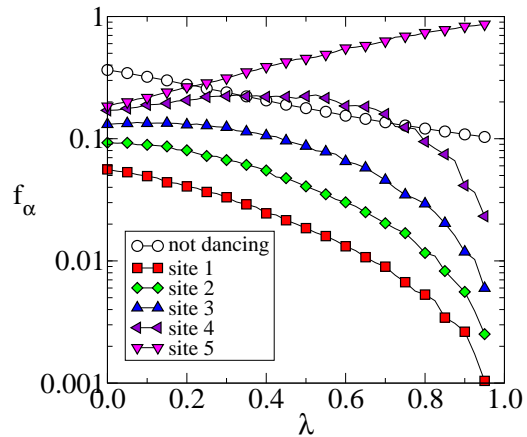}}
\caption{Original model by List et al: Fraction $f_\alpha$ of bees dancing for each of the sites, $\alpha=1,2\dots,5$, in the stationary state, along with the fraction of bees who are not dancing, $f_0$, versus $\lambda$ at fixed $\mu=0$. Markers from simulations ($N=200$, run for $1000$ steps, averaged over $10$ runs). Model parameters are $k=5$, $(q_0,q_1,q_2,q_3,q_4,q_5)=(0.1,3,5,7,9,10)$, $\sigma=0.2$, and $\pi_\alpha=0.05$ for all $\alpha=1,\dots,5$.}
\label{fig:figg1}
\end{figure}

The dynamics as defined above is easily implemented in a computer
simulation, and LES have presented an extensive numerical analysis of
the model. For completeness we present some simulation results in
Fig. \ref{fig:figg1}, where we show the fraction of bees dancing for
the different sites in a situation where there are five potential
nest-sites, and with strong independence ($1-\mu=1$), and varying
interdependence. As seen in the figure, the best nest-site tends to be
the most populated one on average, but convergence to the best
nest-site is strong only at moderate to high interdependence, but not
so much at low values of $\lambda$.
\subsection{Simplified model}
 
If one introduces state variables $x_i(t)=(s_i(t),d_i(t))$, as LES do,
 then the above dynamics defines a Markov chain in the state space
 spanned by the $\{x_i\}$, $i=1,\dots,N$, and a master equation
 approach is feasible in principle. However, as we will discuss
 in the next section, the model can actually be reduced first,
 without affecting its key features and behavior much, to yield a dynamics
 which is Markovian in the space of the $\{s_i(t)\}$ alone. This
 simplifies the analysis considerably, and as we will show next, the
 `memory' variables $d_i(t)$ can be dispensed with entirely.

The simplified model operates purely in the space of the $\{s_i(t)\}$,
i.e. the state of the system at any given time is fully determined by
the variables $s_1(t),\dots,s_N(t)$. In essence the simplification
consists in the following step: if in the original model a bee begins
to dance for a given site at a given time $t$, the duration $d_i(t)$
of this dance is determined at the beginning of the dance, and then,
up to rounding to a near integer, at time step $t+d_i(t)$ the bee
stops dancing for this given site. For $\mu<1$, the duration of the
dance depends on the quality of the site danced for, and is also
subject to some uncertainty as explained above. But crucially, the
duration of the dance is set at the beginning of the dance, the end of
the dance then occurs at a deterministic moment in time, once the
duration has been chosen. The duration of the dance may have a
stochastic element to it as explained above, but its value is fixed
and known when the dance starts. In the reduced model we replace this
with a random process, i.e. a bee may start dancing for a given site
$\alpha$ at time $t$, and then at each subsequent time step the bee
ceases dancing with a certain probability $r_\alpha$ and continues
dancing with probability $1-r_\alpha$. The rate $r_\alpha$ with which
the dance is stopped at each step may here depend on the quality of
the site, for example a high-quality site $\alpha$ will have a lower
value of $r_\alpha$ than a site $\beta$ of a lesser quality. The
probability that a dance for a given site $\alpha$ lasts for precisely
$\ell$ time-steps after the dance has begun is then given by
$(1-r_\alpha)^{\ell-1}r_\alpha$, i.e. it follows a geometric distribution of mean $1/r_\alpha$. This may be slightly unrealistic in
the sense that the probability of stopping after exactly $\ell$ rounds
is a monotonically decreasing function of $\ell$, whereas the
distribution of dance durations used by LES has a well-defined
maximum, but this drawback is, in our view, largely outweighed by the
analytical simplifications it brings with it. As we will show below this
simplification does not affect the qualitative behavior of the model
much. A further restriction of the reduced model is the fact that
there is no clear analogue of the uncertainty parameter $\sigma$ in
the present setup, i.e. once $r_\alpha$ has been chosen, both the mean
and variance of the distribution of durations of dances for site
$\alpha$ are fixed. We will discuss possibilities to include such
uncertainty towards the end of this paper.

The precise algorithm of the simplified model is the following, the definitions and conventions regarding the $\{\pi_\alpha\}$ and the model parameters $\lambda$ and $\mu$ are the same as in the original model:
\begin{enumerate}
\item[1.] At time $t=0$ initialize the $\{s_i(t=0)\}$ for all $i$, for example set all bees in a non-dancing state, or initialize the $s_i(t)$ at random.
\item[2.] At time $t$ compute the fraction $f_\alpha(t)\in[0,1]$ of bees dancing for site $\alpha$ at time $t$ ($\alpha=0,\dots,k$), and compute $p_\alpha(t+1)=(1-\lambda)\pi_\alpha+\lambda f_\alpha(t)$. Note that $0\leq p_\alpha(t+1)\leq 1$ for all $\alpha$, and $\sum_{\alpha=0}^k p_\alpha(t+1)=1$ by definition. 
\item[3.] In order to update the states of the bees distinguish between bees that are not dancing at time $t$ and those which are dancing at time $t$. Update all bees in parallel dynamics, i.e. for all $i=1,\dots,N$ perform the following:
\begin{itemize}
\item[3a)] If bee $i$ is dancing for site $\alpha$ at time $t$ (i.e. if $s_i(t)=\alpha>0$), then with probability $r_\alpha$ set $s_i(t+1)=0$ (bee stops dancing), and with probability $1-r_\alpha$ set $s_i(t+1)=\alpha$ (bee keeps dancing for site $\alpha$).

\item[3b)] If, at time $t$, bee $i$ is not dancing ($s_i(t)=0$), then with probability $p_\alpha(t+1)$ set $s_i(t+1)=\alpha$, $\alpha=0,1,\dots,k$, i.e. with probability $p_1(t+1)$ the bee starts dancing for site $\alpha=1$, with probability $p_2(t+1)$ she starts dancing for site $\alpha=2$ and so on. With probability $p_0(t+1)$ she will remain in the passive state $\alpha=0$.
\end{itemize}
\item[4.] Iterate, i.e. go to 2.
\end{enumerate}
It remains to specify the rates with which dances are terminated, i.e. to define the $\{r_\alpha\}$. We here choose
\be\label{eq:rr}
r_\alpha=q_0\left[\frac{\mu}{K}+\frac{1-\mu}{q_\alpha}\right], ~~~~\alpha=1,\dots,k,
\ee
where $K=q_k$ as in \cite{list}. For any $\mu<1$ the rate $r_\alpha$ is a decreasing function of $q_\alpha$, i.e. the higher a site's quality, the lower the rate at which dances for it are stopped. The pre-factor $q_0$ (which needs to be non-zero) here ensures that $0<r_\alpha\leq 1$ and also that the $\{r_\alpha\}$ do not depend on the overall scale of the nest-site qualities. Equivalently, $q_0$, together with the rate at which dances are commenced, defines the time-scale of the model. A different scaling of time could be chosen by using $\gamma q_0$ as a pre-factor in (\ref{eq:rr}), with $0<\gamma<1$. We note that for $\mu=0$ (maximal independence) one has $r_\alpha=q_0/q_\alpha$, i.e. $r_\alpha$ is (up to the pre-factor) given by the inverse quality of the nest-site $\alpha$, so that the mean duration of dances is proportional to the nest-site quality. For $\mu=1$ (no independent assessment of the nest-site quality), one has $r_\alpha=q_0/K$ irrespective of $\alpha\in\{1,\dots,k\}$, so that the mean duration of dances for a given site does not depend on its quality at all. It is here worth pointing out that if the initial state is chosen to be the all-passive state, then for $\lambda=1$ no bee ever starts dancing.  An analogous effect is present in the original model by LES. We will therefore generally exclude the case of $\lambda=1$ from the analysis, and choose $\lambda$ large, but strictly smaller than unity, if we want to model strong interdependence. Alternatively one can choose random initial conditions.

\section{Analytical solution and test against simulations}\label{sec:ana}
\begin{figure}[t]
\centerline{\includegraphics[width=0.4\textwidth]{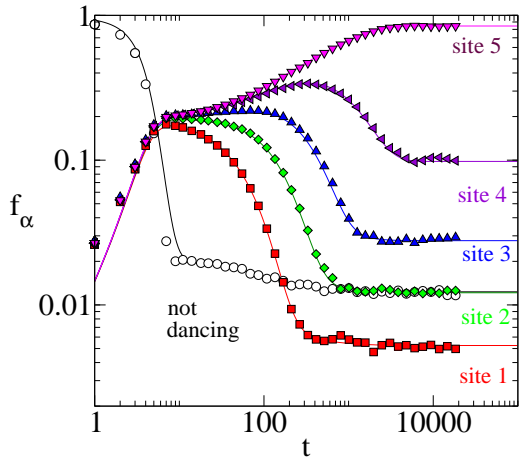}}
\caption{Temporal decision dynamics of the group of bees at $\mu=0, \lambda=0.8$. We show the fraction $f_\alpha(t)$ of bees dancing for each of the sites, $\alpha=1,2,\dots,5$, along with the fraction of bees who are not dancing. Solid lines are from the deterministic dynamics, Eqs. (\ref{eq:mf}), markers from simulations started from an all-passive state ($N=400$ bees, averaged over $250$ runs).}
\label{fig:figg2}
\end{figure}
\begin{figure}[t]
\vspace{2em}
\centerline{\includegraphics[width=0.4\textwidth]{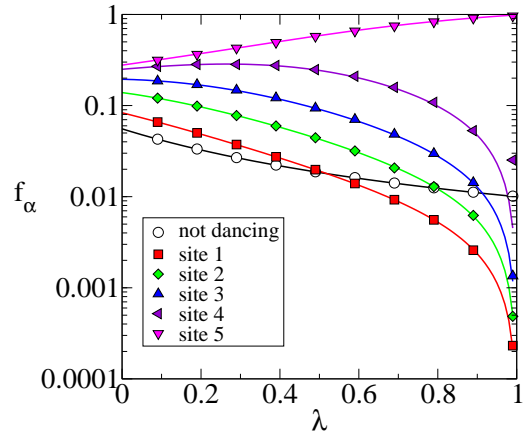}}
\caption{Fraction $f_\alpha^*$ of bees dancing for each of the sites, $\alpha=1,2\dots,5$, in the long-run, along with the fraction of bees who are not dancing, versus $\lambda$ at fixed $\mu=0$. Solid lines are from the deterministic dynamics, markers from simulations ($N=400$, run for $20000$ steps, averaged over $50$ runs).}
\label{fig:figg3}
\end{figure}
\subsection{Master equation and dynamics in the deterministic limit}
In order to analyze the stochastic process defined by the dynamics of the model, we will write $n_\alpha(t)$ for the number of bees dancing for each of the sites at time $t$. We here explicitly include the state $\alpha=0$, i.e. bees not dancing at time $t$. One then has $\sum_{\alpha=0}^k n_\alpha(t)=N$ for all times $t$. For later convenience we introduce the vector $\bn(t)=(n_0(t),n_1(t),\dots,n_k(t))$. The state of the system at time $t$ is therefore fully determined by $\bn(t)$. The reduced model defines a Markov process for $\bn(t)$, described by the following master equation (see textbooks such as \cite{vankampen} or \cite{gardiner} for details on the master equation formalism)
 \BE
\frac{d}{dt}P_\bn &=&\sum_{\alpha=1}^k P_{\bn-\bee_\alpha+\bee_0}T^+_\alpha(\bn-\bee_\alpha+\bee_0)-P_\bn\sum_{\alpha=1}^kT^+_\alpha(\bn) \nonumber\\
&&\hspace{-2em}+\sum_{\alpha=1}^k P_{\bn+\bee_\alpha-\bee_0} T^-_{\alpha}(\bn+\bee_\alpha-\bee_0)-P_\bn\sum_{\alpha=1}^k T^-_\alpha(\bn)\nonumber\\ \label{eq:master}
\EE
for the probability $P_\bn(t)$ of finding the system in state $\bn$ at time $t$. $T^+_\alpha(\bn)$ here stands for the probability that a randomly chosen bee is inactive and starts dancing for site $\alpha\in\{1,\dots,k\}$ at time $t$, given that the system is in state $\bn$. Similarly, $T^-_\alpha(\bn)$ is the probability for a randomly chosen individual to be dancing for site $\alpha$ and for it to cease dancing in the subsequent time step. We write $\bee_\alpha$, $\alpha=0,1,\dots,k$ for the $k+1$ unit vectors, i.e. one has $(\bee_\alpha)_\beta=\delta_{\alpha\beta}$ for $\alpha,\beta\in\{0,1,\dots,k\}$. The transition rates are given by
\be
T_\alpha^+(\bn)=\frac{n_0}{N}\left[(1-\lambda)\pi_\alpha+\lambda \frac{n_\alpha}{N}\right]
\ee
and
\be
T_\alpha^-(\bn)=\frac{n_{\alpha}}{N}\underbrace{q_0\left[\frac{\mu}{K}+\frac{1-\mu}{q_\alpha} \right]}_{=r_\alpha}
\ee
respectively, where $\alpha=1,\dots,k$. Technically speaking, this master equation is more appropriate for so-called random sequential updating, in which the states of the $N$ bees are updated one-by-one (and each update would then take a time-step of $1/N$, so that each bee is updated on average once per unit time), see \cite{vankampen,risken,gardiner} for details. The pre-factors $n_\alpha/N$ in the rates $T_\alpha^\pm$ here in fact represent the probabilities that a randomly chosen bee is in state $\alpha$. We have verified in numerical simulations that there are no significant differences between the stationary states of the model with parallel and sequential updating respectively, and as we will see below theoretical predictions from our master equation approach agree well with simulations using parallel update rules.

Introducing continuous variables $f_\alpha=n_\alpha/N$ for $\alpha=0,1,\dots,k$ it is then straightforward to derive a set of deterministic ordinary differential equations in the limit of an infinite system size, $N\to\infty$. This may for example be done by considering
\be
\avg{\bn(t)}=\sum_\bn \bn P_\bn(t)
\ee
and noting that
\BE
\frac{d\avg{\bn(t)}}{dt}&=&\sum_{\alpha=1}^k \left[(\bee_\alpha-\bee_0)T^+_\alpha(\avg{\bn})\right]\nonumber\\
&&+\sum_{\alpha=1}^k \left[(\bee_0-\bee_\alpha)T^-_\alpha(\avg{\bn})\right],
\EE
where we have used a deterministic approximation to write $\avg{T^{\pm}_\alpha(\bn)}= T^{\pm}_\alpha(\avg{\bn})$. This gives
\begin{equation}\label{eq:mf}
\dot f_\alpha(t)= (1-\rho(t))\left[(1-\lambda)\pi_\alpha+\lambda f_\alpha(t)\right]-r_\alpha f_\alpha(t),
\end{equation}
 for $\alpha=1,\dots,k$, and where $\rho(t)=\sum_{\alpha=1}^k f_\alpha(t)$. The fraction of bees not dancing at time $t$ is given by $f_0(t)=1-\rho(t)$.

\subsection{Comparison with simulations}
Equations (\ref{eq:mf}) are easily integrated numerically for any fixed choice of the model parameters, and we show a specific example in Fig. \ref{fig:figg2}. We here choose the model parameters as in \cite{list}, specifically we use $k=5$, $(q_0,q_1,...,q_5)=(0.1,3,5,7,9,10)$, $K=q_5$, and $\pi_1=...=\pi_5=0.05$, as well as $\pi_0=0.75$. Our definitions of the transition rates require $q_0>0$ and so we choose a small, but non-zero value for $q_0$.  As mentioned above $q_0$ sets the overall time-scale of the model, and is essentially arbitrary, so that the times indicated in Fig. \ref{fig:figg2} do not necessarily have a direct biological interpretation in terms of e.g. hours, days or the total number of nest inspection flights done by the population. As seen in the figure the theoretical description of Eqs. (\ref{eq:mf}) agrees near perfectly with results from simulations, and reproduces even non-monotonic trajectories faithfully. The stationary state of the system may therefore be determined as a fixed point of Eqs. (\ref{eq:mf}), i.e. as the solution of the $k$ coupled quadratic equations,
\begin{equation}\label{eq:fp}
r_\alpha f_\alpha^*= \left(1-\sum_{\beta=1}^kf_\beta^*\right)\left[(1-\lambda)\pi_\alpha+\lambda f_\alpha^*\right]
\end{equation}
for $\alpha=1,\dots,k$, and where we have introduced the notation $f_\alpha^*$ to denote the fixed-points of (\ref{eq:mf}). These fixed-point equations may in principle be solved numerically, or alternatively, one integrates Eqs. (\ref{eq:mf}) to asymptotic times. We have here followed the latter approach. It is interesting to note that Eqs. (\ref{eq:fp}) can be re-arranged to give
\be\label{eq:order}
f_\alpha^*=\left[\frac{r_\alpha}{1-\rho^*}-\lambda\right]^{-1}\bigg(1-\lambda\bigg)\pi_\alpha, ~~~\alpha=1,\dots,k.
\ee
Given that all sites have the same a-priori probability of being found, i.e. that $\pi_\alpha$ does not depend on $\alpha\in\{1,\dots,k\}$, and that secondly the rates of dances being terminated decrease with an increasing quality of the sites, $r_\alpha<r_\beta$ for $\alpha>\beta$, Eqs. (\ref{eq:order}) reveal immediately, that a site is the more populated on the deterministic level, the higher its quality, i.e. $f_1^*<f_2^*<\dots<f_k^*$, in-line with the results of LES. We also note that for $\lambda=1$ the only possible fixed point is one at which $f_\alpha^*=0$ for all $\alpha=1,\dots,k$.

To test our theoretical predictions regarding the stationary state we show the asymptotic values of $f_0^*,f_1^*,...,f_5^*$ as a function of the model parameter $\lambda$ in Fig. \ref{fig:figg3}, and again simulations confirm that our mean-field description accurately reproduces the behavior of the agent-based model in a wide range of model parameters. Figure \ref{fig:figg3} also reveals that the behavior of the simplified model is very similar to that of the original model by LES (see Hypothesis 1 and Figs. 1-3 of \cite{list}): if the bees assess the quality of the sites independently before dancing for a given site, i.e. if $\mu=0$, then the best site is retrieved successfully over a wide range of values for the parameter $\lambda$. In-line with the findings of \cite{list}, an increased amount of interdependence between the bees (high values of $\lambda$) increases the chances of finding the best nest site, as indicated by the proportion of bees, $f_5^*$, dancing for the best site $\alpha=5$.

\begin{figure}[t]
\vspace{1em}
\centerline{\includegraphics[width=0.4\textwidth]{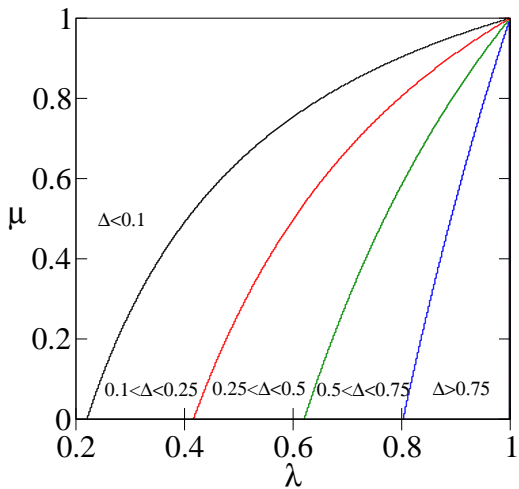}}\vspace{2em}
\vspace{1em}
\caption{Quality of the retrieval of the best site, as measured by $\Delta$ (see main text), throughout the parameter space. High values of $\Delta$ indicate a high accuracy in identifying the best nest site. Results are from an integration of Eqs. (\ref{eq:mf}), using a first-order Euler-forward scheme with a time-stepping of $dt=0.1$, integrated up to time $t=10^5$.}
\label{fig:figg4}
\end{figure}

Having established the correctness of our theoretical approach, we can now push the analysis further, and assess the decision-dynamics of the bees throughout the parameter space defined by $\lambda$ and $\mu$. To this end we have integrated Eqs. (\ref{eq:mf}) to large times ($t=10^5$) for approximately $1.6\cdot 10^5$ different combinations of $\lambda$ and $\mu$, equally spaced in the unit square, and have computed the difference $\Delta=f_5^*-f_4^*$ between the fraction of bees dancing for the best site and the fraction of bees dancing for the second-best site. $\Delta$ therefore ranges in the interval $[0,1]$, and high values of $\Delta$ indicate that the best site ($\alpha=5$) is identified with a high accuracy by the population of bees. Results are shown in Fig. \ref{fig:figg4}, and confirm that a reliable convergence to the best site can only be achieved in the presence of both, a sufficient degree of independence in the assessment of nest-sites and interdependence between the different individuals. This is a confirmation of the second main hypothesis of \cite{list}. The best retrieval of the optimal site is found at low values of $\mu$ (high independence) and large $\lambda$ (high interdependence). In order to check for convergence we have also analyzed data from an integration of Eqs. (\ref{eq:mf}) only up to $t=10^4$. The location of the different lines in Fig. \ref{fig:figg4} remains unchanged at this lower time, so that we conclude that the stationary state has already been reached much earlier than at $t=10^5$.

\subsection{Size-limitations of best-site retrieval}    
\begin{figure}[t]
\vspace{1em}
\centerline{\includegraphics[width=0.4\textwidth]{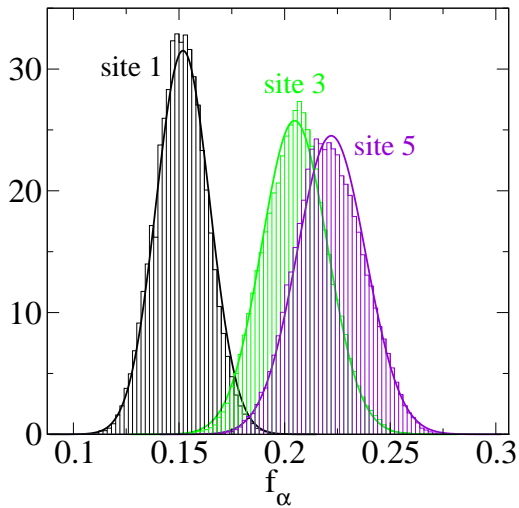}}
\vspace{1em}
\caption{The population levels $n_\alpha/N$ are stochastic quantities for finite systems, we here show their distributions for sites $\alpha=1$, $\alpha=3$ and $\alpha=5$ for fixed model parameters $\mu=0.95, \lambda=0.4$. The system-size is $N=2500$. Measurements in simulations are taken after an equilibration period of $10000$ time-steps, and results are averaged over $20$ independent runs. Solid lines are the theoretical predictions derived from Eqs. (\ref{eq:dotsigma}), i.e. they are Gaussian distributions of mean $f_\alpha^*$ and with variances $\overline\Lambda_{\alpha\alpha}$.}
\label{fig:figg5a}
\end{figure}

\begin{figure}[t]
\vspace{0em}
\centerline{\includegraphics[width=0.4\textwidth]{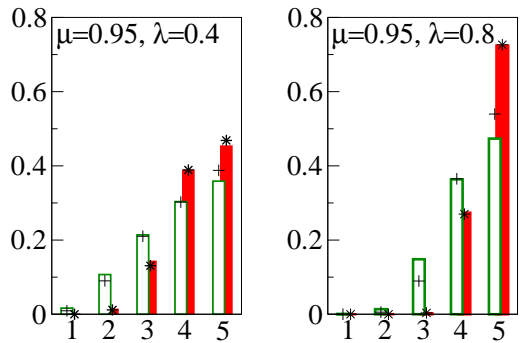}}
\vspace{2em}
\caption{Probabilities for the different sites $\alpha$ to be the most-populated one. Filled bars show results from simulations for $N=2500$, averaged over $20$ samples, open green bars are for $N=400$ (averaged over $50$ samples); the measurement period starts after $10000$ time-steps in the simulations. The stars and crosses show the estimates from the analytical theory in the first order of the system-size expansion, as described in the text. }
\label{fig:figg5}
\end{figure}

The deterministic dynamics defined by Eqs. (\ref{eq:mf}) approaches a
stationary state in which the proportion of bees populating the best
site is higher than that of any other site. The description in terms
of these deterministic equations is however only valid in the formal
limit of an infinite population of bees, i.e. if $N\to\infty$. In
small populations stochastic effects may instead dominate the decision
making, and as observed by LES convergence to a sub-optimal state may
occur. To characterize these finite-size effects one may look at a
system at a fixed size $N$ and at fixed model parameters $\lambda$ and
$\mu$. At any given time, the state of the system is described by the
(re-scaled) population vector $(n_1(t)/N,n_2(t)/N,\dots,n_k(t)/N)$,
and the dominating choice at any given time may well be different from
the optimal site, i.e. it may well be that the most-populate site at
time $t$ is not actually the best site $\alpha=k$. This is the case
whenever there is an index $\alpha\in\{1,\dots,k-1\}$ such that
$n_\alpha(t)>n_k(t)$. It is therefore instructive to look at the joint
and marginal distributions of the values
$(n_1(t)/N,n_2(t)/N,\dots,n_k(t)/N)$ sampled over time and from
different simulation runs. We first address the marginals, and the
resulting histograms are shown for a particular choice for $\lambda$
and $\mu$ and for a relatively large system ($N=2500$) in
Fig. \ref{fig:figg5a}. For optical convenience we suppress the results
for the second and fourth sites respectively. As seen in the figure
there may be a considerable overlap between the distributions
corresponding to the optimal site and the sub-optimal ones, so that
the population of bees may well misidentify the best site with a
significant probability. Since the variance of the distributions shown
in Fig. \ref{fig:figg5a} scales as $N^{-1}$ by virtue of the central
limit theorem, this effect will generally be the less pronounced the
larger the swarm size. For large swarms the individual population
levels fluctuate less, resulting in sharper peaks, and a lesser chance
of favoring a sub-optimal state. This is a special case of what is
known as `Condorcet's jury theorem' (see \cite{list} and references
therein), and the effect is illustrated in Fig. \ref{fig:figg5}, where
we plot the frequencies with which the different sites are favored in
simulations at fixed model parameters and system sizes. More
precisely, in each simulation run, we check for each time step which
site is the most populated at this moment in time, and from this
construct the histograms shown in the figure, indicating how
frequently the different sites come up as the most-populated one (the
`site' $\alpha=0$ (passive bees) is here not taken into account). We
take a simple majority as sufficient criterion for a consensus,
although this is admittedly a rather weak definition (termed `weak
consensus criterion' by LES), stronger conditions are discussed in
\cite{list}. As seen in the figure site $\alpha=5$ comes up with the
highest rate for the model parameters chosen in the figure, but
there is also a considerable probability for the sub-optimal sites to
have the most bees dancing for them. From our simulations we conclude
that this effect is relevant mostly at small independence ($\mu$ close
to unity), and tends to become stronger the smaller the
interdependence parameter $\lambda$, and/or the smaller the swarm.

These finite-size stochastic effects can actually be understood analytically, based on an approach using the so-called van Kampen expansion in the inverse system size. In the remainder of this section we will describe some of the intermediate steps, but will not go through the derivation in all detail, as the necessary mathematics is readily available in the literature \cite{vankampen}. As a starting point we separate off stochastic fluctuations from the deterministic solution $\boldeff(t)$ of Eqs. (\ref{eq:mf}):
\be
\frac{\bn(t)}{N}=\boldeff(t)+\frac{1}{\sqrt{N}}\bxi(t),
\ee
anticipating that the magnitude of deviations from the deterministic system scales as $N^{-1/2}$. One then proceeds as described in \cite{vankampen}, and expands the master equation (\ref{eq:master}) in powers of $N^{-1/2}$. In lowest order one recovers the deterministic dynamics of Eqs. (\ref{eq:mf}), and in next-to-leading order one has
\be\label{eq:langevin}
\dot\bxi=\mathbb{J}\bxi+\boldeta
\ee
in the stationary regime, where a fixed-point of the mean-field trajectory has been assumed. The matrix $\mathbb{J}$ is the $k\times k$ Jacobian of the deterministic dynamics at this fixed point, i.e. one has
\be
J_{\alpha\beta}=-\left[(1-\lambda)\pi_\alpha+\lambda f_\alpha^*\right]+\left[\lambda(1-\rho^*)-r_\alpha\right]\delta_{\alpha\beta} 
\ee 
for $\alpha,\beta\in\{1,\dots,k\}$. The quantity $\boldeta$ in (\ref{eq:langevin}) is Gaussian white noise of zero mean and with the following diagonal covariance matrix between components
\be
\avg{\eta_\alpha(t)\eta_\beta(t')}=\delta_{\alpha\beta}\Gamma_{\alpha\beta}\delta(t-t'),
\ee
where
\be\label{eq:gamma}
\Gamma_{\alpha\alpha}=(1-\rho^*)\left[(1-\lambda)\pi_\alpha+\lambda f_\alpha^*\right]+r_\alpha f_\alpha^*.
\ee
These results can also be read off from the general expressions given for example in \cite{boland}. Equation (\ref{eq:langevin}) describes an multi-component Ornstein-Uhlenbeck process \cite{gardiner}, and the co-variance matrix of the random variables $\xi_\alpha(t)$ hence has a relatively simple time-evolution. In particular if we define $\Lambda_{\alpha\beta}(t)=\avg{\xi_\alpha(t)\xi_\beta(t)}$, then one has \cite{risken}
\be\label{eq:dotsigma}
\dot \Lambda_{\alpha\beta}=\sum_{\gamma=1}^k\left[J_{\alpha\gamma}\Lambda_{\gamma\beta}+J_{\beta\gamma}\Lambda_{\gamma\alpha}\right]+\Gamma_{\alpha\beta},
\ee
for $\alpha,\beta\in\{1,\dots,k\}$. Denoting the fixed-point of these equations by $\overline{\Lambda}$ one has asymptotically
\BE\label{eq:gauss}
&&P\left(\frac{\bn}{N}=\bx^*+N^{-1/2}\bxi\right)\nonumber\\
&=&{\cal N}\exp\left(-\frac{1}{2}\sum_{\alpha\beta}\xi_\alpha(\overline{\Lambda}^{-1})_{\alpha\beta}\xi_\beta\right)
\EE
for the distribution of the population vector $\bn$. ${\cal N}$ is a normalization factor. Within our approximation (consisting in the truncation of the van-Kampen expansion after the sub-leading order) this distribution is Gaussian, with a co-variance matrix $\overline{\Lambda}$. It is here appropriate to point out that, within this approximation, the expression in Eq. (\ref{eq:gauss}) characterizes the statistics of the system at finite sizes fully, as both the variances of the individual components of $\bn$ and their correlations are taken into account. This allows one to estimate mathematically how often each of the sites will be the `winner' of the quorum decision, for example, using the weak consensus criterion of LES.

In order to generate these semi-analytical estimates characterizing the finite-size effects of the system, we have first integrated the master equation to asymptotic times to extract the fixed-point vector $\boldeff^*$. From this one computes the Jacobian $\mathbb{J}$, and the above matrix $\Gamma$. Integrating Eqs. (\ref{eq:dotsigma}) then gives $\overline{\Lambda}$. A comparison of these findings against simulations is shown in Fig. \ref{fig:figg5a}, and as seen in the figure reasonably good agreement is achieved so that we conclude that our theoretical approach gives accurate estimates for the variances of the $\{n_\alpha\}$, at least for the model parameters used in Fig. \ref{fig:figg5a}. The matrix $\overline{\Lambda}$ is then inverted, and subsequently results are inserted into Eq. (\ref{eq:gauss}), and numerical integration of the resulting multi-variate Gaussian distribution then produces analytical estimates for the probabilities for the different sites $\alpha\in\{1,\dots,k\}$ to be the most-populated one. These semi-analytical results are compared with simulations in Fig. \ref{fig:figg5}, and the agreement is found to be reasonable, especially at large system sizes. We would here like to stress that we have not been able to obtain agreement of the level shown in Figs. \ref{fig:figg5a} and \ref{fig:figg5} for all values of the model parameters. In particular if $\lambda$ is large, deviations can be significant, even for the largest system-sizes we have tried (up to $10^4$). We attribute this to remaining finite-size corrections (truncating the van Kampen expansion might not be justified in these circumstances), to equilibration effects, or to numerical inaccuracies in integrating the $5$-dimensional Gaussian distribution (\ref{eq:gauss}). Also note that the Gaussian approximation becomes inaccurate if any of the $f_\alpha^*$ are near the upper or lower limits of the interval $[0,1]$. In simulations no $n_\alpha/N$ can ever come out negative or exceed unity, these boundary effects are not captured by the first-order approximation of the van Kampen expansion. As a final note in this section, we point out that the stronger consensus criterion discussed in \cite{list} can be studied based on Eq. (\ref{eq:gauss}) as well, as this equation provides a Gaussian estimate for the joint distribution of all $k$ population levels. We have however not tried to do so here.

\section{Spatial model}\label{sec:space}
\subsection{Model definitions and simulation results}
In this section we consider a spatial extension of the simplified
model. To this end we place the dancing or resting bees on a square
lattice of lateral extension $L$, and apply periodic boundary
conditions for simplicity. The total system is composed of $N=L\times
L$ agents. The update dynamics is as in the simplified model described
above, the only difference is that any given non-dancing bee is only
affected by bees in its direct neighborhood when it decides to start
dancing for a given site, and not by the entire
population. Specifically, the above fractions $f_\alpha(t)$, of bees
dancing for site $\alpha$, now carry a dependence on space, i.e. we
introduce $f_\alpha(\bx,t)$ as the fraction of bees on nearest
neighbor lattice sites of $\bx$ dancing for site $\alpha$. More
precisely we have
$f_\alpha(\bx,t)=\frac{1}{4}\sum_{\by\sim\bx}\delta_{\alpha,s_\by(t)}$, where the
sum over $\by$ extends over the four nearest neighbors of $\bx$. A
bee located at lattice site $\bx$ then takes into account the
$f_\alpha(\bx,t), ~\alpha=1,\dots,k$ when it considers to commence a
dance. The rate with which it starts dancing for site $\alpha$ is then given by $p_\alpha(\bx,t)=(1-\lambda)\pi_\alpha+\lambda f_\alpha(\bx,t)$. Apart from this modification the dynamics proceeds exactly as in the non-spatial model.
\begin{figure*}[t]
\vspace{4em}
\centerline{\hspace{-3em}\includegraphics[width=0.75\textwidth]{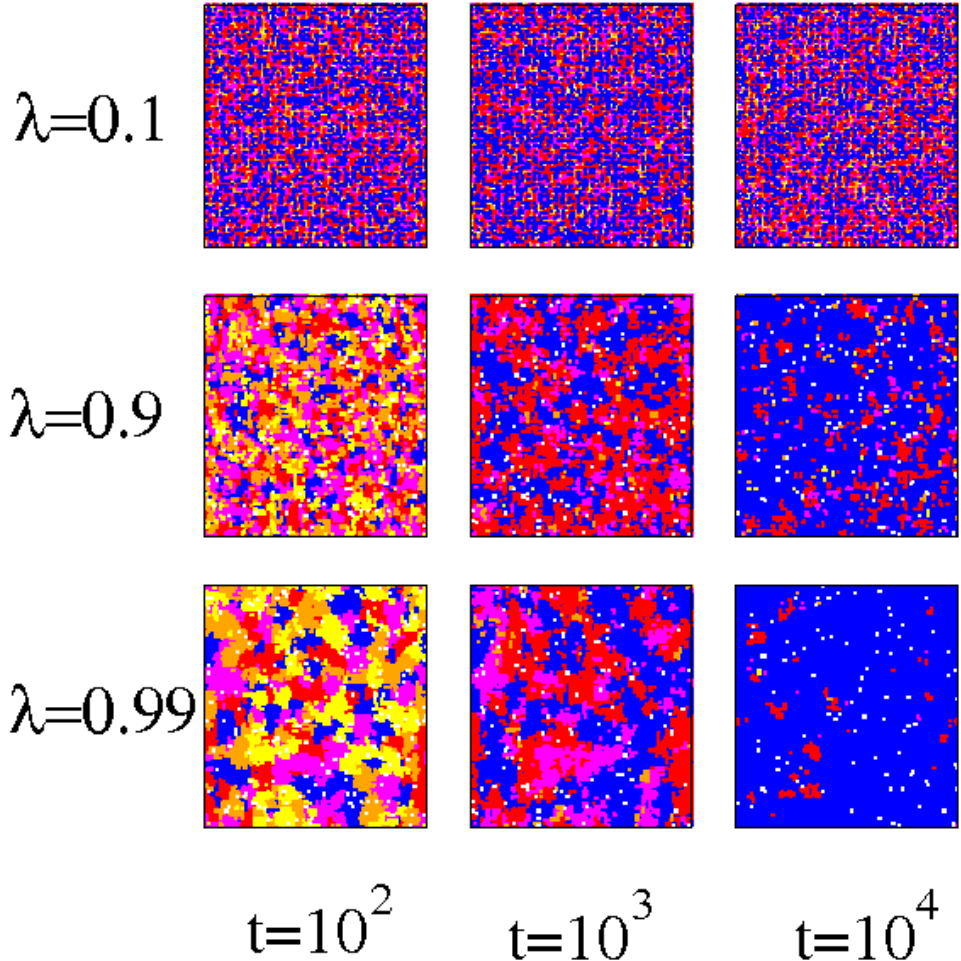}}
\vspace{3em}
\caption{Pattern formation and coarsening dynamics in the spatial model at $\mu=0$. The panels show the configuration of the system at $\lambda=0.1,0.9$ and $\lambda=0.99$ respectively, and at different times of the simulation runs (i.e. after $t=10^2,10^3,10^4$ sweeps). Simulations are performed on a square lattice with $L\times L=100^2$ agents,  with periodic boundary conditions, and started from an all-passive initial condition. The different colors indicate different states of the individual bees, with white squares corresponding to bees who are not dancing, and filled squares to bees dancing for a particular site as indicated by the different colors. Lighter colors correspond to the states with low quality, and the black (blue on-line) areas to bees dancing for the best site. The optical impression may sometimes not reflect the actual proportions of bees dancing for the actual sites, accurately, e.g. in the top row we have $27\%$, $31\%$ and $31\%$ of bees dancing for site $5$, in the bottom row these numbers are $24\%$, $52\%$ and $97\%$ at the three different times respectively.}
\label{fig:fig7}
\end{figure*}

One should point out that we are here departing from a realistic modelling of the decision making of bees, as space in our model is isotropic and invariant against translations. No special point in space is designated as the entrance into the hive (where incoming bees would arrive when they return from site-inspection flights). However, as discussed in the conclusions, the dynamics proposed by LES can be understood as a more general model for decision making, applicable also in contexts different from the nest-site choice of bees, so that it is worthwhile investigating the spatial extension discussed in the present section. 

Results from simulations are shown in Fig. \ref{fig:fig7}. When the simulations are initiated, all bees are passive, i.e. not dancing. As seen in the figure non-trivial large-scale patterns emerge for sufficiently large interdependence, see especially the configurations corresponding to $\lambda=0.9$ and $\lambda=0.99$ at intermediate times $t\approx 100$. Inspecting the system at later times (right-most snapshots in Fig. \ref{fig:fig7}) shows that these patterns are of a transient nature only, and a coarsening dynamics to a broad consensus on the best nest-site emerges in the spatial system.  Interestingly, transient structures do depend on the interdependence parameter $\lambda$: strong interdependence promotes large domains of agreement on one of the different nest sites, whereas at small $\lambda$ no such large-scale areas of consensus are found. We should here also point out that these patterns do seem to depend on the initial conditions chosen, e.g. if we initiate the system from a state where all bees are active and dancing for randomly chosen sites, then the formation of domains seems to be far less pronounced.

\subsection{Analytical approaches}
In order to make further progress towards an analytical description of
the spatial system, we here extend the above master equation approach,
and study the following spatial analogue of the mean-field dynamics
(\ref{eq:mf}) of the non-spatial system, \BE\label{eq:mfspace} \dot
\varphi_\alpha(\bx,t)&=&
\bigg(1-\rho(\bx,t)\bigg)\bigg[(1-\lambda)\pi_\alpha\nonumber\\
&&+\lambda (\Delta \varphi_\alpha
(\bx,t)+\varphi_\alpha(\bx,t))\bigg]\nonumber \\ &&-r_\alpha
\varphi_\alpha(\bx,t) \EE for $\alpha=1,\dots,k$. Here
$\varphi_\alpha(\bx,t)$ is a continuous-valued field defined on the
lattice points of the system, and indicates the probability of finding
the agent at site $\bx$ in state $\alpha$. The $\varphi_\alpha(\bx,t)$
are hence coarse-grained order parameters, but no continuum limit in
space is implied. Note that Eqs. (\ref{eq:mfspace}) are coupled
through $\rho(\bx,t)=\sum_{\alpha=1}^k\varphi_\alpha(\bx,t)$. The
quantity $\Delta\varphi_\alpha(\bx,t)$ is the lattice Laplacian,
i.e. we have
$\Delta\varphi_\alpha(\bx,t)=\left(\frac{1}{4}\sum_{\by\sim\bx}\varphi_\alpha(\by,t)\right)-\varphi_\alpha(\bx,t)$. The term $\Delta \varphi_\alpha
(\bx,t)+\varphi_\alpha(\bx,t)$ in the above expression of the right-hand-side of Eqs. (\ref{eq:mfspace}) is thus equal to $\frac{1}{4}\sum_{\by\sim\bx} \varphi_\alpha(\by,t)$, and describes the total density of individuals in state $\alpha$ in the neighbourhood of $\bx$. 

\begin{figure}[t]
\vspace{1em}
\centerline{\hspace{-3em}\includegraphics[width=0.4\textwidth]{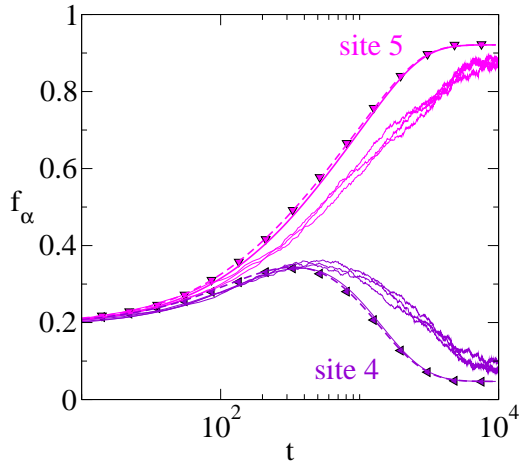}}
\vspace{1em}
\caption{Fraction of bees dancing for the best and second-best sites respectively (upper curves resp. lower curves). The smooth solid lines show results from an integration of the deterministic dynamics of Eqs. (\ref{eq:mfspace}), the dashed lines are from an integration of the non-spatial deterministic dynamics given by Eqs. (\ref{eq:mf}). Noisy lines show three independent simulation runs of the spatial agent-based model for $N=100^2$ agents. Markers are from simulations of the non-spatial model ($N=10^4$, average over $10$ samples). Model parameters are $\mu=0, \lambda=0.9$, in the initial state all bees are active and dance for randomly chosen sites.}
\label{fig:figg8}
\end{figure}

\begin{figure*}[t]
\centerline{
\includegraphics[width=0.65\textwidth]{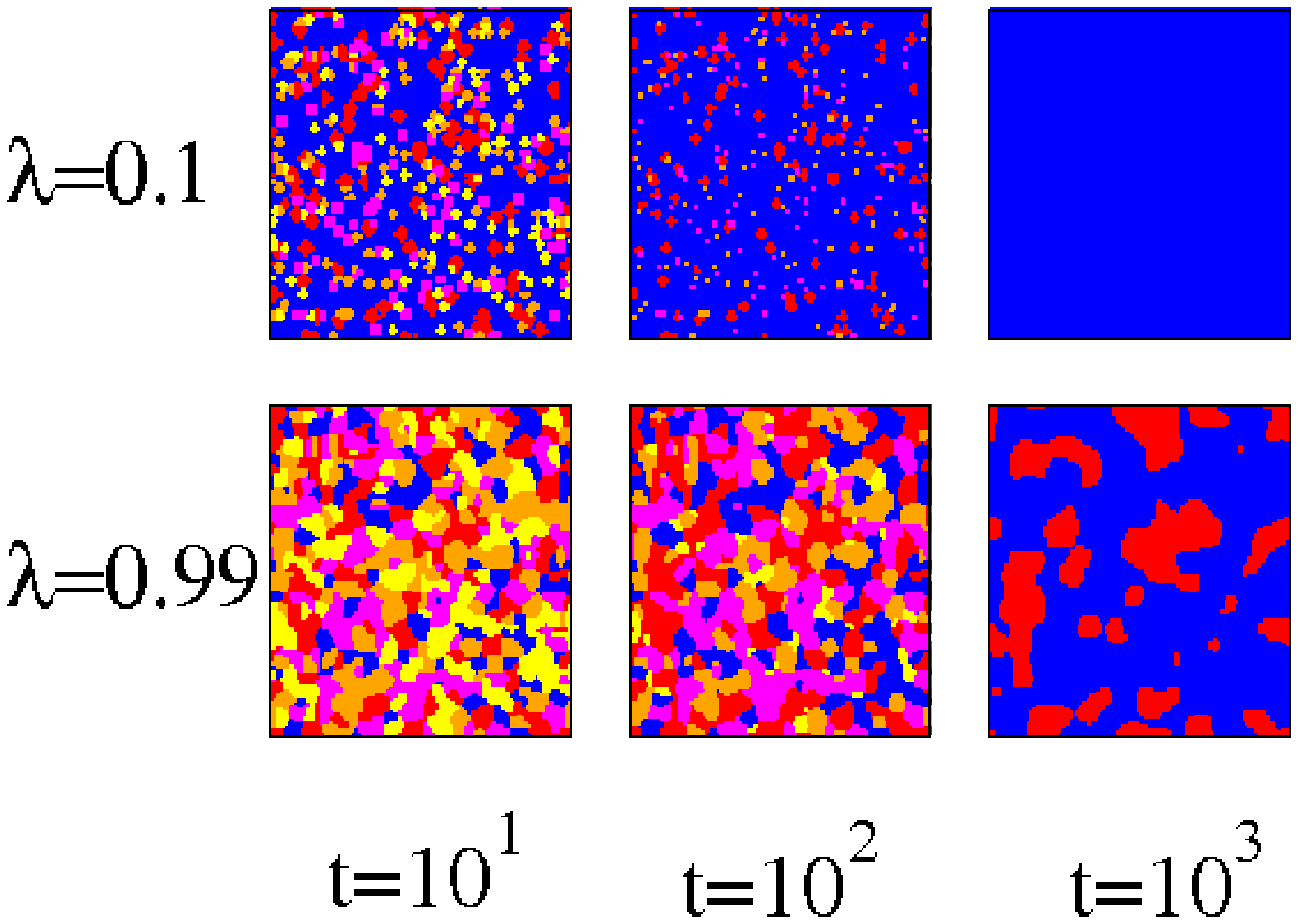}}
\vspace{1em}
\caption{Field configurations generated from a numerical integration of Eqs. (\ref{eq:mfspace}) at $\mu=0$, started from an initial condition in which 95$\%$ of all bees are passive, and the remaining randomly chosen $5\%$ of bees dance for randomly chosen sites (i.e. at each lattice point draw one $\alpha\in\{1,\dots,k\}$ and set $\varphi_\alpha(\bx,t=0)=1$, and $\varphi_\beta(\bx,t)=0$, for all $\beta\neq\alpha$). Integration is performed using a first-order Euler-forward scheme with a time stepping of $dt=0.1$. The different colors indicate the dominant state, i.e. at each lattice site $\bx$ and each moment in time we identify $\alpha$ such that $\varphi_\alpha=\mbox{max}\{\varphi_1(\bx,t),...,\varphi_k(\bx,t)\}$. Color coding is as in Fig. \ref{fig:fig7}. The size of the lattice is $100\times 100$.}
\label{fig:fig9}
\end{figure*}

Results
from a numerical integration of these equations
are shown in Fig. \ref{fig:figg8}, for a case in which random initial
conditions are chosen to introduce spatial inhomogeneity. We find that
the prediction from the spatial and non-spatial deterministic dynamics
do not differ much, and that Eqs. (\ref{eq:mfspace}) capture the
qualitative behavior of the dynamics, but that no quantitative
agreement is achieved.  The dynamics on the lattice appears to
converge more slowly than in the well-mixed system, and than predicted
by Eqs. (\ref{eq:mfspace}). These deviations point to the observation
that while neglecting fluctuations was justified in the well-mixed
non-spatial system, stochasticity is relevant in the two-dimensional
system. In order to examine the spatial structures one obtains from
Eqs. (\ref{eq:mfspace}) we show a number of snapshots in
Fig. \ref{fig:fig9}. It should be noted that these were taken from
random initial conditions, with $5\%$ of the bees active, as explained
in the figure caption. Homogeneous initial conditions in the
noise-less equations (\ref{eq:mfspace}) do not lead to any pattern
formation, as the field trivially remains homogeneous for all
times. Fig. \ref{fig:fig9} shows domain formation similar to that in
Fig. \ref{fig:fig7}, but the extent to which the patterns of the
agent-based model are reproduced by the deterministic dynamics of
Eqs. (\ref{eq:mfspace}) is at present still undetermined, and will be
left for future work. Integrating Eqs. (\ref{eq:mfspace}) at
$\lambda=0.9$ for example (not shown) yields results very similar to
those at $\lambda=0.99$, whereas in the agent-based model systematic
differences are found between those two cases, as seen in
Fig. \ref{fig:fig7}. To understand the relation between the
agent-based model, and the effective dynamics in terms of continuous
fields a detailed analysis of e.g. correlation lengths and other
statistical features of the spatio-temporal patterns will be
required. Based on the observations reported in Fig. \ref{fig:figg8}
it may well be the case that a more accurate description will make it
necessary to take into account fluctuations around the deterministic
dynamics. We have not attempted to pursue this further, but based on
existing work in voter models and related spatial agent-based systems
(see e.g. \cite{chate}, \cite{frey} or \cite{dallasta} and references
therein), one may consider the following stochastic extension
\BE\label{eq:mfspacenoise} \dot \varphi_\alpha(\bx,t)&=&
\bigg(1-\rho(\bx,t)\bigg)\bigg[(1-\lambda)\pi_\alpha\nonumber\\
&&+\lambda (\Delta \varphi_\alpha
(\bx,t)+\varphi_\alpha(\bx,t))\bigg]\nonumber \\ &&-r_\alpha
\varphi_\alpha(\bx,t)+\zeta_\alpha(\bx,t), \EE where
$\zeta_\alpha(\bx,t)$ is Gaussian noise of zero mean and with
correlations \be\label{eq:spacecov0}
\avg{\zeta_\alpha(\bx,t)\zeta_\beta(\bx',t')}=\delta_{\alpha\beta}B_{\alpha\alpha}\delta(\bx-\bx')\delta(t-t'),
\ee where \BE
B_{\alpha\alpha}&=&\bigg(1-\rho(\bx,t)\bigg)\bigg[(1-\lambda)\pi_\alpha\nonumber\\
&&+\lambda (\Delta \varphi_\alpha
(\bx,t)+\varphi_\alpha(\bx,t))\bigg]\nonumber \\ &&+r_\alpha
\varphi_\alpha(\bx,t).\label{eq:spacecov} \EE This covariance matrix
is an analogue of (\ref{eq:gamma}) and can be obtained following
\cite{dallasta}. Due to the multiplicative noise, it is difficult to
integrate equations
(\ref{eq:mfspacenoise},\ref{eq:spacecov0},\ref{eq:spacecov})
numerically, so that we do not present any further results here, but
only mention that one approach might be to extend recently developed
methods \cite{moro,chate} to the case of multiple fields. We would
like to stress that
(\ref{eq:mfspace},\ref{eq:mfspacenoise},\ref{eq:spacecov0},\ref{eq:spacecov})
are not to be understood as a rigorously valid description of the
spatial system. We feel that further more detailed considerations of
the continuous description of the model are beyond the scope of the
present paper, but would like to mention that several approaches could
be considered in future work: explicit diffusion (particle exchange
between neighbouring sites) could be taken into account, and a
continuum limit could be derived along the lines of \cite{frey}, using
a Kramers-Moyal expansion (see \cite{gardiner}, and also
\cite{traulsen}). Alternatively systematic expansions for spatial
models with a large number of individuals per lattice site can be
found in \cite{lugo}, where a van Kampen system-size expansion is
used \cite{vankampen}.

\section{Conclusions}\label{sec:concl}
The model of the nest-site choice dynamics in swarms of honeybees by
LES is a beautiful example of a biologically inspired agent-based
model, which can be solved with techniques originally developed in
statistical physics.  As we have shown, only modest modifications to
the dynamics proposed by LES are required to make the model analytically
tractable. The reduced model is Markovian in a relatively simple
configuration space, and we have used a master equation approach to
derive a deterministic continuous-time description in terms of a set
of coupled differential equations. These can be integrated
numerically, and alternatively their fixed points can be obtained as a
the solution of a small set of quadratic equations. The reduction may limit the degree of realism in the model, but as we have
shown the simpler model reproduces most, if not all features of the
original dynamics. In particular it confirms that in order for the
population of bees to accurately identify the best site for a future
nest, both independence in the assessment of the quality of the nest
sites and interdependence between the bees is necessary. Based on our
analytical solutions we are able to characterize the stationary state
of the system in the entire parameter space spanned by the variables
$\lambda$ and $\mu$ representing the degree of interdependence and the
degree of independence respectively, without the need to perform extensive agent-based simulations. While the analytical description is mostly on the deterministic level, a systematic approach to finite-size effects is feasible based on an expansion in the inverse system size. This allows one, within the limitations discussed above, to obtain estimates for the frequency with which finite swarms of bees will reach a consensus for each of the potential nest-sites, including the sub-optimal ones.

We have also extended the model to a spatial arrangement on a simple square lattice, and have demonstrated in computer simulations that the interacting agent model gives rise to non-trivial transient patterns before a coarsening dynamics towards a consensus on the best nest-site sets in. These patterns are observed only at moderate to high interdependence of the bees. Finally, we have discussed a simple set of reaction-diffusion like equation derived from the master equation of the spatial system. These equations capture some of the characteristics of the spatial patterns formed in the agent-based model, but further work is required to determine whether they faithfully capture statistical features of these patterns. It is important to note that we have not taken into account explicit diffusion though, as no hopping of agents or particle exchange is considered in our model. A more detailed analysis of the effects of multiplicative noise on the field equations is a further interesting point, which needs a more detailed investigation.

We would finally like to comment on possibilities to account for uncertainties in the quality assessment as LES do in their original model (parametrized by $\sigma$). One way of doing this may be to formally increase the number of sites in our simplified model, i.e. to introduce several `copies' of each site. E.g. for a given site $\alpha$, one could introduce formal copies $\alpha^{(1)}, \alpha^{(2)},...$ and then draw formal quality factors $q_{\alpha^{(1)}},q_{\alpha^{(2)}},\dots$ from some distribution centered around the true quality of site $\alpha$, and with a standard deviation given by the uncertainty parameter $\sigma$. This would mathematically increase the number of `sites' in the model, even though all $\alpha^{(j)}$, $j=1,2,\dots$ would refer to the same real-world site, i.e. the total number of bees dancing for site $\alpha$ would be computed as $n_\alpha^{tot}=\sum_j n_{\alpha^{(j)}}$.

Other points of further research might address the model in spatial arrangements different from a simple two-dimensional regular lattice. One may here for example study the effects of the topology of the network of interactions on the convergence and consensus properties. In this context the model proposed by LES and the simplified dynamics discussed in the present paper could probably be interpreted more generally as a model for the interplay of independence and interdependence in the decision making of groups, where the specific application may not necessarily be limited to the nest-site choice of honeybees. LES give an example towards the end of their paper, relating to the choice of restaurants by humans, and similarly many other situations, where individuals have to make strategic choices, for example in the social sciences and in game theory may potentially be modeled using similar dynamics. The analytical approaches developed in the present paper may here be useful for to characterize the outcome of the decision making processes in such models.


\begin{acknowledgments} 
The author is grateful to Richard Morris for bringing the model by LES to his attention, and acknowledges an RCUK Fellowship (RCUK reference EP/E500048/1).
\end{acknowledgments}


\begin{thebibliography}{10}



\bibitem[List {\em et al.} 2009]{list}
C. List, C. Elsholtz, and T. D. Seeley 2009, Independence and interdependence in collective decision making: an agent-based model of nest-site choice by honeybee swarms. {\em Phil. Trans. Roy. Soc. B.} {\bf 364}, 755. 
\bibitem[van Kampen 1992]{vankampen} N. G. van Kampen, Stochastic processes in physics and chemistry, Elsevier Amsterdam, 1992.
\bibitem[Gardiner 2009]{gardiner} C. Gardiner, Stochastic methods, a handbook for the natural and social sciences, 4th ed., Springer-Verlag Berlin Heidelberg, 2009.
\bibitem[Boland {\em et al.} 2008]{boland} R. P. Boland, T. Galla, and A. J. McKane 2008, How limit cycles and quasi-cycles are related in systems with intrinsic noise. J. Stat. Mech.  (2008) P09001.
\bibitem[Risken 1996]{risken} H. Risken 1996, The Fokker-Planck equations, 2nd ed., Springer-Verlag Berlin Heidelberg, 1996. 
\bibitem[Dall'Asta \& Galla 2008]{dallasta} L. Dall'Asta and T. Galla 2008, Algebraic coarsening in voter models with intermediate states. {\em J. Phys. A: Math. Theor.} {\bf 41} 435003. 
\bibitem[Moro 2004]{moro} E. Moro 2004, Numerical schemes for continuum models of reaction-diffusion systems subject to internal noise. {\em Phys. Rev. E} {\bf 70}, 045102.
\bibitem[Dornic {\em et al.} 2005]{chate} I. Dornic, H. Chat\'e and M. A. Mu\~noz 2005 Integration of Langevin equations with multiplicative noise and the viability of field theories for absorbing phase transitions. {\em Phys. Rev. Lett.} {\bf 94}, 100601.
\bibitem[Reichenbach {\em et al.} 2008]{frey} T. Reichenbach, M. Mobilia and E. Frey 2008, Self-organization of mobile populations in cyclic competition. {\em J. Theor. Biol.} {\bf 254}, 368.
\bibitem[Traulsen {\em et al.} 2006]{traulsen} A. Traulsen, J. C Claussen, C. Hauert 2006 Coevolutionary dynamics in large, but finite populations. {\em Phys. Rev. E} {\bf 74} 011901
\bibitem[Lugo \& McKane 2008]{lugo}C. A. Lugo, A. J. McKane 2008, Quasicycles in a spatial predator-prey model. {\em Phys. Rev. E.} {\bf 78} 051911 






\end{thebibliography}
\end{document}